\documentclass{acmconf}
\usepackage{color}
\usepackage{times}
\usepackage{mathptm}
\usepackage{anysize}
\usepackage{graphicx}
\usepackage{latexsym}
\usepackage{amsfonts}
\usepackage{amsmath}
\usepackage{amssymb}

\input{Qcircuit}

\marginsize{1in}{1in}{1in}{1in}

\newcommand{\braket}[2]{\langle #1 ~|~ #2 \rangle}

\newtheorem{lemma}{Lemma}

\begin{document}
\pagestyle{plain}

\title{Checking Equivalence of Quantum Circuits and States}

\author{George F. Viamontes\footnote{gviamont@atl.lmco.com}\\
 {\em Lockheed Martin Advanced Technology Laboratories, Cherry Hill, NJ 08002}\\ \\
 Igor L. Markov and John P. Hayes\footnote{\{imarkov, jhayes\}@eecs.umich.edu}\\
 {\em Department of EECS, University of Michigan, Ann Arbor, MI 48109-2121}
}

\date{}

\maketitle

\abstract{Quantum computing promises exponential speed-ups for
important simulation and optimization problems.  It also poses new CAD
problems that are similar to, but more challenging, than the related
problems in classical (non-quantum) CAD, such as determining if two
states or circuits are functionally equivalent. While differences in
classical states are easy to detect, quantum states, which are
represented by complex-valued vectors, exhibit subtle differences
leading to several notions of equivalence. This provides flexibility
in optimizing quantum circuits, but leads to difficult new
equivalence-checking issues for simulation and synthesis. We identify
several different equivalence-checking problems and present algorithms
for practical benchmarks, including quantum communication and search
circuits, which are shown to be very fast and robust for hundreds of
qubits.}


\section{Introduction}

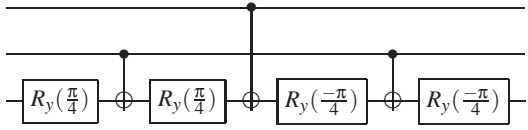
\begin{figure}
\begin{center}
\[
\Qcircuit @C=.7em @R=.1em @!R {
& \qw & \qw & \qw & \ctrl{2} & \qw & \qw & \qw & \qw \\
& \qw & \ctrl{1} & \qw & \qw & \qw & \ctrl{1} & \qw & \qw \\
& \gate{R_y(\frac{\pi}{4})} & \targ & \gate{R_y(\frac{\pi}{4})} & \targ & \gate{R_y(\frac{- \pi}{4})} & \targ & \gate{R_y(\frac{- \pi}{4})} & \qw
}
\]
\parbox{7.5cm}{\caption{\label{fig:margolus}Margolus' circuit is
equivalent up to relative phase to the Toffoli gate, which otherwise
requires six $CNOT$ and eight $1$-qubit gates to implement
\cite{Shende_comm}.}}
\end{center}
\vspace{-6mm}
\end{figure}

Quantum computing (QC) is a recently discovered  alternative to conventional
computer technology  that offers not  only miniaturization, but massive
performance speed-ups for certain tasks \cite{Hey99,Shor1997,Grover1997}
and new levels of protection in secure communications  \cite{BB84, B92}.
Information is stored in particle states and processed 
using quantum-mechanical operations referred to as quantum gates.
The analogue of the classical bit, qubit, has two basic states denoted
$\ket{0}$ and $\ket{1}$, but can also exist in a superposition 
of these two states $\ket{\phi} = \alpha \ket{0} + \beta \ket{1}$,
where $|\alpha|^2+ |\beta|^2=1$.
A composite system consisting of $n$ such qubits requires $2^n$ parameters
(amplitudes) indexed by $n$-bit binary numbers 
$\ket{\Phi}=\Sigma_{i=1}^{2^n} \alpha_i \ket{i}$, 
where $\Sigma |\alpha_i|^2=1$. Quantum gates transform such states
by applying unitary matrices to them.  Measurement of a quantum state
produces classical bits with probabilities dependent on $\alpha_i$.
%
%
%
Combining several gates, as in Figure \ref{fig:margolus},
yields {\em quantum circuits} \cite{NielsenC2000} that compactly
describe more sophisticated transformations that play the role of
quantum algorithms. 

Based on the success of CAD for classical logic circuits, new
algorithms have been proposed for synthesis and simulation of quantum
circuits \cite{BarencoEtAl1995, ShendeEtAl2006, Song2003,
Gottesman1998, AaronsonG2004, ViamontesEtAl2005, Vidal2003}. In
particular, the DAC 2007 paper \cite{MaslovFM07}, describes what
amounts to placement and physical synthesis for quantum circuits ---
``adapting the circuit to particulars of the physical environment
which restricts/complicates the establishment of certain direct
interactions between qubits.'' Another example is given in
\cite[Section 6]{ShendeEtAl2006}.\footnote{For example, in a spin
chain architecture the qubits are laid out in a line, and all CNOT
gates must act only on adjacent (nearest-neighbor) qubits. The work in
\cite{ShendeEtAl2006} shows that such a restriction can be accomodated
by restructuring an existing circuit in such a way that worst-case
circuit sizes grow by no more than nine times.} Traditionally, such
transformations must be verified by equivalence-checking, but the
quantum context is more difficult because qubits and quantum gates may
differ by global and relative phase (defined below), yet be equivalent
upon measurement \cite{NielsenC2000}. To this end, our work is the
first to develop techniques for quantum phase-equivalence checking.

Two quantum states $\ket{\psi}$ and $\ket{\varphi}$ are equivalent
up to {\em global phase} if $\ket{\varphi} = e^{i \theta} \ket{\psi}$,
where $\theta \in \mathbb{R}$. The phase $e^{i \theta}$ will not be observed
upon measurement of either state \cite{NielsenC2000}. By contrast, two
states are equal up to {\em relative phase} if a unitary diagonal matrix can
transform one into the other:

 \begin{equation}
   \label{eq:relative_phase}
   \ket{\varphi} = \mathrm{diag}(e^{i \theta _0}, e^{i \theta_1}, \ldots, e^{i \theta _{N - 1}}) \ket{\psi}.
 \end{equation}

\noindent
The probability amplitudes of the state $U \ket{\psi}$ will in general
differ by more than relative phase from those of $U \ket{\varphi}$,
but the measurement outcomes may be equivalent. 
One can consider a hierarchy in which
exact equivalence implies global-phase equivalence, which implies
relative-phase equivalence, which in turn implies measurement outcome
equivalence. The equivalence checking problem is also extensible to
quantum operators with applications to quantum-circuit synthesis and
verification, which involves computer-aided generation of minimal
quantum circuits with correct functionality.
Extended notions of equivalence create several design
opportunities. For example, the well-known three-qubit Toffoli
gate can be implemented with fewer controlled-NOT (CNOT) and
$1$-qubit gates up to relative phase \cite{BarencoEtAl1995, Song2003}
as shown in Figure \ref{fig:margolus}. The relative-phase differences
can be canceled out if every pair of these gates in the circuit is 
strategically placed
\cite{Song2003}. Since circuit minimization is being pursued for a
number of key quantum arithmetic circuits with many Toffoli gates,
such as modular exponentiation \cite{VanMeter2005, Cuccaro2004,
ShendeEtAl2005, ShendeEtAl2006}, this optimization could reduce the
number of gates even further.

The inner product and matrix product may be used to determine such
equivalences, but in this work, we present new decision-diagram (DD)
algorithms to accomplish the task more efficiently. In particular, we
make use of the quantum information decision diagram (QuIDD)
\cite{ViamontesEtAl2003, ViamontesEtAl2005}, a datastructure with
unique properties that are exploited to solve this problem
asymptotically faster in practical cases.

Empirical results confirm the algorithms' effectiveness and show that
the improvements are more significant for the operators than for the
states. Interestingly, solving the equivalence problems for the
benchmarks considered requires significantly less time than creating
the DD representations, which indicates that such problems can be
reasonably solved in practice using quantum-circuit CAD tools.

The structure of this work is as follows. Section \ref{sec:background}
provides a review of the QuIDD datastructure. Section
\ref{sec:global_phase} describes both linear-algebraic and QuIDD
algorithms for checking global-phase equivalence of states and
operators. Section \ref{sec:relative_phase} covers relative-phase
equivalence checking algorithms. Sections \ref{sec:global_phase} and
\ref{sec:relative_phase} also contain empirical studies comparing the
algorithms' performance on various benchmarks. Lastly, conclusions and
a summary of computational complexity results for all algorithms are
provided in Section \ref{sec:conclusions}.

\section{Background}
\label{sec:background}

\begin{figure*}[htb]
  \begin{center}
    \begin{tabular}{c|cc}
      \hspace{30mm}\begin{tabular}{c}
	\raisebox{25ex}[0pt]{\small $\mspace{-5mu}\begin{array}{rl}
	    R_0 R_1 & \\
	    \begin{array}{c}
	      0\ 0 \\
	      0\ 1 \\
	      1\ 0 \\
	      1\ 1
	    \end{array} &
	    \mspace{-25mu}\left[\begin{array}{c}
		0.707107 \\
		-0.707107 \\
		0.707107 \\
		-0.707107
	      \end{array}\right]
	  \end{array}$} \\
	\raisebox{19ex}[0pt]{\small $\updownarrow$} \\
	\hspace{-3.7em}\raisebox{2ex}[0pt]{\includegraphics[width=3.7cm]{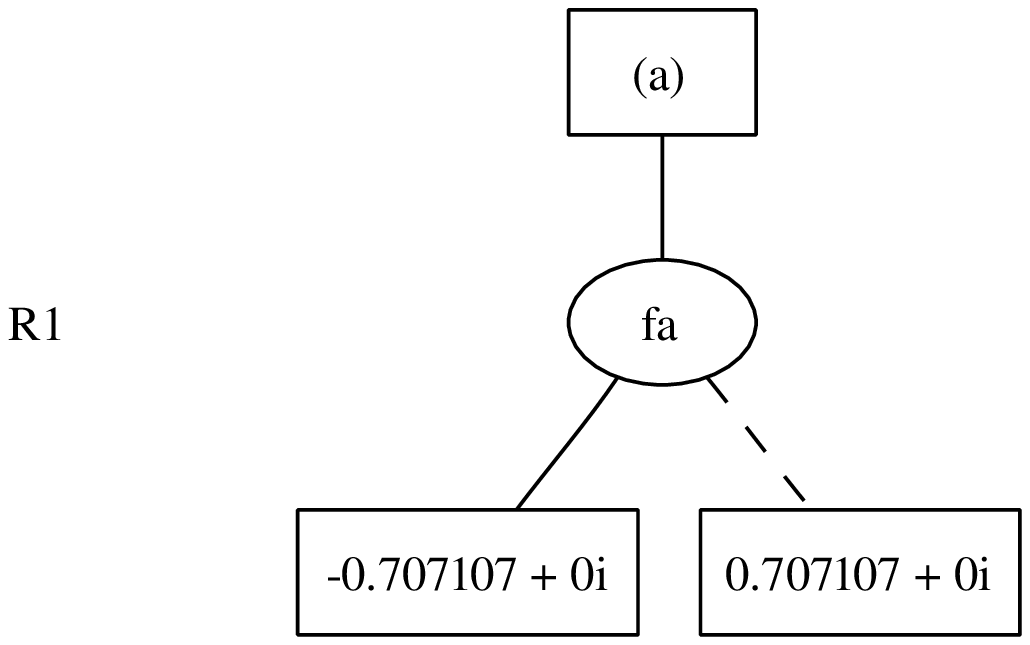}}
      \end{tabular} &
      \includegraphics[width=3.5cm]{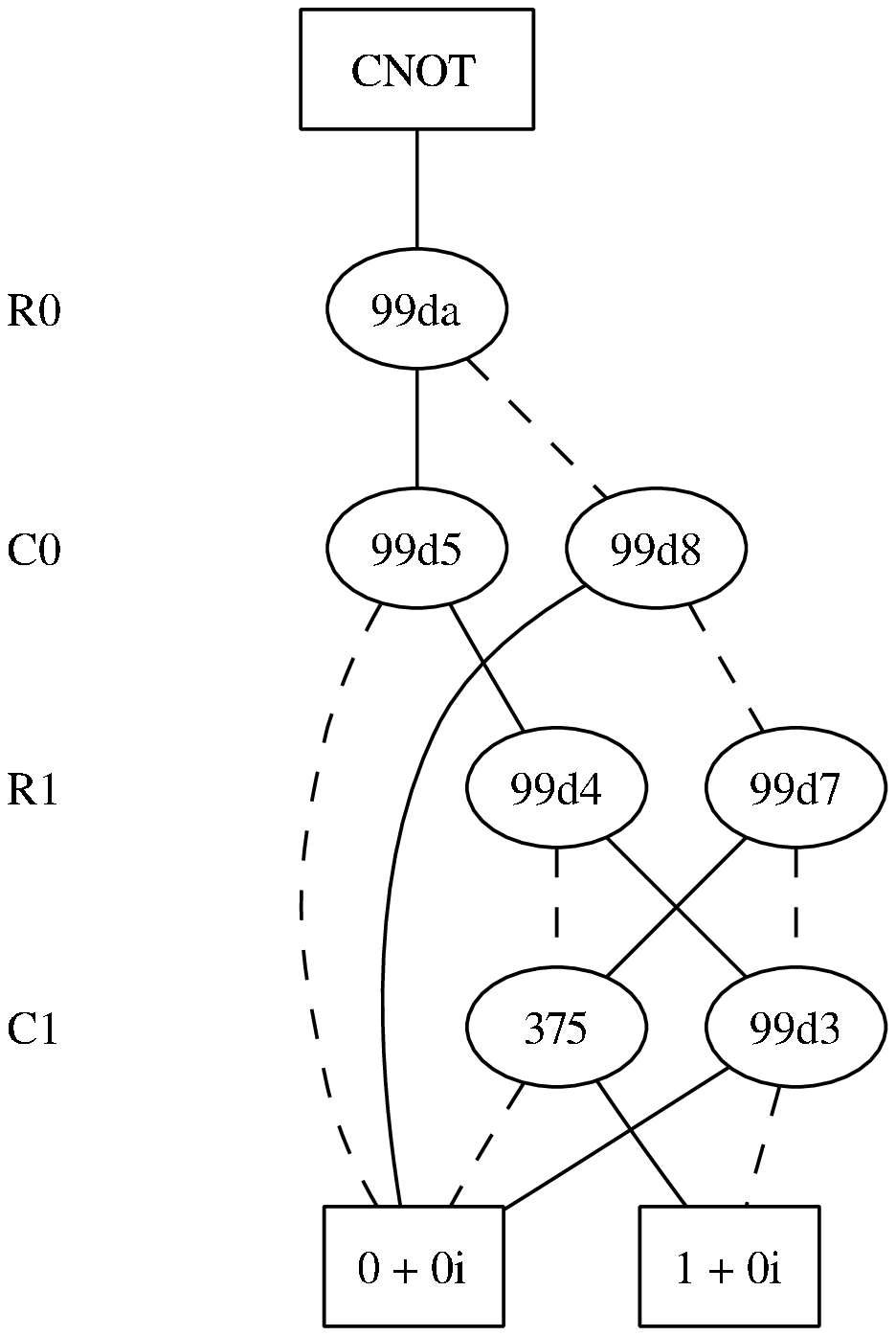} &
      \raisebox{15ex}[0pt]{\small $\mspace{-15mu}\leftrightarrow \mspace{-10mu} \begin{array}{rc}
	& C_0 C_1 \\
	R_0 R_1 & \mspace{-25mu} 00 \mspace{9mu} 01 \mspace{9mu} 10 \mspace{9mu} 11 \\
	\begin{array}{c}
	  0\ 0 \\
	  0\ 1 \\
	  1\ 0 \\
	  1\ 1
	\end{array} &
	\mspace{-25mu}\left[\begin{array}{cccc}
	    1 & 0 & 0 & 0 \\
	    0 & 1 & 0 & 0 \\
	    0 & 0 & 0 & 1 \\
	    0 & 0 & 1 & 0
	  \end{array}\right]
      \end{array}$} \\
      \hspace{20mm}(a) & \multicolumn{2}{c}{(b)} \\
    \end{tabular}
    \parbox{16cm}{\caption{\label{fig:quidds}Sample QuIDDs of (a) a
    $2$-qubit equal superposition with relative phases and (b) the
    $CNOT$ operator. Each internal node (circle) is unique and depends
    on a variable listed to the left (dashed (solid) edge is $0$ ($1$)
    assignment). Internal node labels are unique hexadecimal
    identifiers based on each node's memory address. Terminal nodes
    (squares) contain complex values.}}
  \end{center}
\vspace{-3mm}
\end{figure*}

The QuIDD is a variant of the reduced ordered binary decision diagram
(ROBDD or BDD) datastructure \cite{Bryant86} applied to quantum
circuit simulation \cite{ViamontesEtAl2003, ViamontesEtAl2005}. Like
other DD variants, it has all of the key properties of BDDs as well as
a few other application-specific attributes (see Figure
\ref{fig:quidds} for examples).

\begin{itemize}
  
  \addtolength{\baselineskip}{-0.8mm}

  {\item It is a directed acyclic graph with internal nodes whose
  edges represent assignments to binary variables}

  {\item The leaf or terminal nodes contain complex values}
  
  {\item Each path from the root to a terminal node is a functional
  mapping of row and column indices to complex-valued matrix elements
  ($f : \{0, 1\}^n \to \mathbb{C}$)}

  {\item Nodes are unique and shared, meaning that any nodes $v$ and
  $v'$ with isomorphic subgraphs do not exist}

  {\item Variables whose values do not affect the function output for
  a particular path (not in the {\em support}) are absent}

  {\item Binary row ($R_i$) and column ($C_i$) index variables have
  evaluation order $R_0 \prec C_0 \prec \ldots R_{n - 1} \prec C_{n -
  1}$}

\end{itemize}

The algorithms which manipulate DDs are just as important as the
properties of the DDs. In particular, the $\mathbf{Apply}$ algorithm
(see Figure \ref{fig:apply}) performs recursive traversals on DD
operands to build new DDs using any desired unary or binary function
\cite{Bryant86}. Although originally intended for digital logic
operations, $\mathbf{Apply}$ has been extended to linear-algebraic
operations such as matrix addition and multiplication \cite{Bahar97,
Clarke96}, as well as quantum-mechanical operations such as
measurement and partial trace \cite{ViamontesEtAl2003,
ViamontesEtAl2005}. The runtime and memory complexity of
$\mathbf{Apply}$ is $O(|A||B|)$, where $|A|$ and $|B|$ are the sizes
in number of internal and terminal nodes of the DDs $A$ and $B$,
respectively \cite{Bryant86}.\footnote{The runtime and memory
complexity of the unary version acting on one DD $A$ is $O(|A|)$
\cite{Bryant86}.} Thus, the complexity of DD-based algorithms is tied
to the compression achieved by the datastructure. These complexity
bounds are important for analyzing many of the algorithms presented in
this work.

\begin{figure}[tb]
  \begin{center}
    \begin{tabular}{c}
      \framebox[7cm][l]{
      \parbox{8cm}{
	$\begin{array}{l}
	  \mathbf{Apply}(A, B, b\_op)\ \{ \\
	  \hspace{1em}\mathbf{if}\ (Is\_Constant(A)\ \mathbf{and}\ Is\_Constant(B))\ \{ \\
	  \hspace{2em}\mathbf{return}\ New\_Terminal(b\_op(Value(A),\\
	  \hspace{3em}Value(B))); \\
	  \hspace{1em}\} \\
	  \hspace{1em}\mathbf{if}\ (Table\_Lookup(R, b\_op, A, B))\ \mathbf{return}\ R; \\
	  \hspace{1em}v\ =\ Top\_Var(A, B); \\
	  \hspace{1em}T\ =\ \mathbf{Apply}(A_v, B_v, b\_op); \\
	  \hspace{1em}E\ =\ \mathbf{Apply}(A_{v'}, B_{v'}, b\_op); \\
	  \hspace{1em}R\ =\ ITE(v, T, E); \\
	  \hspace{1em}Table\_Insert(R, b\_op, A, B); \\
	  \hspace{1em}\mathbf{return}\ R; \\
	  \}
	\end{array}
	$
      }}
    \end{tabular}
    \parbox{7cm}{\caption{\label{fig:apply}The $\mathbf{Apply}$
algorithm. $Top\_Var$ returns the smaller variable index from $A$ or
$B$, while $ITE$ creates a new internal node with children $T$ and
$E$.}}
  \end{center}
  \vspace{-6mm}
\end{figure}

Another important aspect of $\mathbf{Apply}$ is that it utilizes a
cache of internal nodes and binary operators ($Table\_Lookup$ and
$Table\_Insert$) to ensure that the new DD being created obeys the DD
uniqueness properties. Maintaining these properties makes many DDs
such as QuIDDs canonical, meaning that two different DDs do not
implement the same function. Thus, exact equivalence checking is
trivial with canonical DDs and may be performed in $O(1)$ time by
comparing the root nodes, a technique which has been long exploited in
the classical domain \cite{tsunami}. Quantum state and operator
equivalence is less trivial as we show.

\section{Checking Equivalence up to Global Phase}
\label{sec:global_phase}

This section describes algorithms that check global-phase equivalence
of two quantum states or operators. The first two algorithms are known
QuIDD-based linear-algebraic operations, while the remaining
algorithms are the new ones that exploit DD properties explicitly. The
section concludes with experiments comparing all algorithms.


\subsection{Inner Product Check}
\label{sec:inner_product}

Since the quantum-circuit formalism models an arbitrary quantum state
$\ket{\psi}$ as a unit vector, then the inner product
$\braket{\psi}{\psi} = 1$. In the case of a global-phase difference
between two states $\ket{\psi}$ and $\ket{\varphi}$, the inner product
is the global-phase factor, $\braket{\varphi}{\psi} = e^{i \theta}
\braket{\psi}{\psi} = e^{i \theta}$. Since $|e^{i \theta}| = 1$ for
any $\theta$, checking if the complex modulus of the inner product is
$1$ suffices to check global-phase equivalence for states.




Although the inner product may be computed using explicit arrays, a
QuIDD-based implementation is easily derived. The complex-conjugate
transpose and matrix product with QuIDD operands have been previously
defined \cite{ViamontesEtAl2003}. Thus, the algorithm computes the
complex-conjugate transpose of $A$ and multiplies the result with
$B$. The complexity of this algorithm is given by the following lemma.


\begin{lemma}
  \label{lemma:inner_prod}
  Consider state QuIDDs $A$ and $B$ with sizes $|A|$ and $|B|$,
  respectively, in nodes. Computing the global-phase difference via
  the inner product uses $O(|A||B|)$ time and memory.
\end{lemma}

\noindent
{\bf Proof.} Computing the complex-conjugate transpose of $A$ requires
$O(|A|)$ time and memory since it is a unary call to $\mathbf{Apply}$
\cite{ViamontesEtAl2003}. Matrix multiplication of two ADDs of sizes
$|A|$ and $|B|$ requires $O((|A||B|)^2)$ time and memory
\cite{Bahar97}. However, this bound is loose for an inner product
because only a single dot product must be performed. In this case, the
ADD matrix multiplication algorithm reduces to a single call of $C =
\mathbf{Apply}(A, B, *)$ followed by $D = \mathbf{Apply}(C, +)$
\cite{Bahar97}. $D$ is a single terminal node containing the
global-phase factor if $|value(D)| = 1$. $\mathbf{Apply}(A, B, *)$ and
$\mathbf{Apply}(C, +)$ are computed in $O(|A||B|)$ time and memory
\cite{Bryant86}, while $|value(D)|$ is computed in $O(1)$ time and
memory. \hfill $\Box$

\subsection{Matrix Product}
\label{sec:matrix_product}

The matrix product of two operators can be used for global-phase
equivalence checking. In particular, since all quantum operators are
unitary, the adjoint of each operator is its inverse. Thus, if two
operators $U$ and $V$ differ by a global phase, then $U V^{\dagger} =
e^{i \theta} I$.

With QuIDDs for $U$ and $V$, computing $V^{\dagger}$ requires $O(|V|)$
time and memory \cite{ViamontesEtAl2003}. Computing $W = U
V^{\dagger}$ requires $O((|U||V|)^2)$ time and memory
\cite{Bahar97}. To check if $W = e^{i \theta} I$, any terminal value
$t$ is chosen from $W$, and scalar division is performed as $W' =
\mathbf{Apply}(W, t, /)$, which takes $O((|U||V|)^2)$ time and
memory. Canonicity ensures that checking if $W' = I$ requires only
$O(1)$ time and memory. If $W' = I$, then $t$ is the global-phase
factor.

\subsection{Node-Count Check}
\label{sec:node_count_check}

The previous algorithms merely translate linear-algebraic operations
to QuIDDs, but exploiting the following QuIDD property leads to faster
checks.


\begin{lemma}
  \label{lemma:scalar_iso}
  The QuIDD $A' = \mathbf{Apply}(A, c, *)$, where $c \in \mathbb{C}$
  and $c \ne 0$, is isomorphic to $A$, hence $|A'| = |A|$.
\end{lemma}

\noindent
{\bf Proof.} In creating $A'$, $\mathbf{Apply}$ expands all of the
internal nodes of $A$ since $c$ is a scalar, and the new terminals are
the terminals of $A$ multiplied by $c$. All terminal values $t_i$ of
$A$ are unique by definition of a QuIDD
\cite{ViamontesEtAl2003}. Thus, $c t_i \ne c t_j$ for all $i, j$ such
that $i \ne j$. As a result, the number of terminals in $A'$ is the
same as in $A$. \hfill $\Box$


Lemma \ref{lemma:scalar_iso} states that two QuIDD states or operators
that differ by a non-zero scalar, such as a global-phase factor, have
the same number of nodes. Thus, equal node counts in QuIDDs are a
necessary but not sufficient condition for global-phase
equivalence. To see why it is not sufficient, consider two state
vectors $\ket{\psi}$ and $\ket{\varphi}$ with elements $w_j$ and
$v_k$, respectively, where $j, k = 0, 1, \ldots N - 1$. If some $w_j =
v_k = 0$ such that $j \ne k$, then $\ket{\varphi} \ne e^{i \theta}
\ket{\psi}$. The QuIDD representations of these states can in general
have the same node counts. Despite this drawback, the node-count check
requires only $O(1)$ time since $\mathbf{Apply}$ is easily augmented
to recursively sum the number of nodes as a QuIDD is created.


\subsection{Recursive Check}
\label{sec:global_rec}

Lemma \ref{lemma:scalar_iso} implies that a QuIDD-based algorithm can
implement a sufficient condition for global-phase equivalence by
accounting for terminal value differences. The pseudo code for such an
algorithm ($\mathbf{GPRC}$) is presented in Figure
\ref{fig:global_rec}.

\begin{figure}[tb]
  \begin{center}
    \footnotesize
    \begin{tabular}{c}
      \framebox[7.5cm][l]{
      \parbox{8cm}{
	$\begin{array}{l}
	  \mathbf{GPRC}(A, B, gp, have\_gp)\ \{ \\
	  \hspace{1em}\mathbf{if}\ (Is\_Constant(A)\ \mathbf{and}\ Is\_Constant(B))\ \{ \\
	  \hspace{2em}\mathbf{if}\ (Value(B) == 0)\ \ \mathbf{return}\ (Value(A) == 0); \\
	  \hspace{2em}ngp = Value(A)/Value(B); \\
	  \hspace{2em}\mathbf{if}\ (\mathbf{sqrt}(\mathbf{real}(ngp)*\mathbf{real}(ngp) +\\
	  \hspace{3em}\mathbf{imag}(ngp)*\mathbf{imag}(ngp))\ !=\ 1) \\
	  \hspace{4em}\mathbf{return}\ \mathbf{false}; \\
	  \hspace{2em}\mathbf{if}\ (!have\_gp)\ \{ \\
	  \hspace{3em}gp = ngp; \\
	  \hspace{3em}have\_gp = \mathbf{true}; \\
	  \hspace{2em}\} \\
	  \hspace{2em}\mathbf{return}\ (ngp == gp); \\
	  \hspace{1em}\} \\
	  \hspace{1em}\mathbf{if}\ ((Is\_Constant(A)\ \mathbf{and}\ !Is\_Constant(B)) \\
	  \hspace{1em}\ \ \ \mathbf{or}\ (!Is\_Constant(A)\ \mathbf{and}\ Is\_Constant(B))) \\
	  \hspace{3em}\mathbf{return}\ \mathbf{false}; \\
	  \hspace{1em}\mathbf{if}\ (Var(A) != Var(B))\ \mathbf{return}\ \mathbf{false}; \\
	  \hspace{1em}\mathbf{return}\ (\mathbf{GPRC}(Then(A), Then(B), gp, have\_gp) \\
	  \hspace{1em}\ \ \mathbf{and}\ \mathbf{GPRC}(Else(A), Else(B), gp, have\_gp)); \\
	  \}
\end{array}
$
}}
\end{tabular}
\parbox{7cm}{\caption{\label{fig:global_rec}Recursive global-phase
equivalence check.}}
\end{center}
\vspace{-7mm}
\end{figure}

$\mathbf{GPRC}$ returns $\mathbf{true}$ if two QuIDDs $A$ and $B$
differ by global phase and $\mathbf{false}$ otherwise. $gp$ and
$have\_gp$ are global variables containing the global-phase factor and
a flag signifying whether or not a terminal node has been reached,
respectively. $gp$ is defined only if $\mathbf{true}$ is returned.

The first conditional block of $\mathbf{GPRC}$ deals with terminal
values. The potential global-phase factor $ngp$ is computed after
handling division by $0$. If $|ngp| \ne 1$ or if $ngp \ne gp$ when
$gp$ has been set,then the two QuIDDs do not differ by a global
phase. Next, the condition specified by Lemma \ref{lemma:scalar_iso}
is addressed. If the node of $A$ depends on a different row or column
variable than the node of $B$, then $A$ and $B$ are not isomorphic and
thus cannot differ by global phase. Finally, $\mathbf{GPRC}$ is called
recursively, and the results of these calls are combined via the
logical $AND$ operation.

Early termination occurs when isomorphism is violated or more than one
phase difference is computed. In the worst case, both QuIDDs are
isomorphic and all nodes are visisted, but the last terminal visited
in each QuIDD will not be equal up to global phase. Thus, the overall
runtime and memory complexity of $\mathbf{GPRC}$ for states or
operators is $O(|A| + |B|)$. Also, the node-count check can be run
before $\mathbf{GPRC}$ to quickly eliminate many nonequivalences.

\subsection{Empirical Results for Global-Phase\\Equivalence Algorithms}
\label{sec:gp_results}

\begin{figure}
  \begin{center}
    \[
    \Qcircuit @C=1em @R=.4em @!R {
      & \lstick{\ket{0}} & \gate{H} & \gate{H} & \multigate{6}{CPS} & \gate{H} & \ctrl{1} & \qw & \qw & \qw \\
      & \lstick{\ket{0}} & \gate{H} & \gate{H} & \ghost{CPS} & \gate{H} & \ctrl{1} & \qw & \qw & \qw \\
      & \lstick{\ket{0}} & \gate{H} & \gate{H} & \ghost{CPS} & \gate{H} & \ctrl{1} & \qw & \qw & \qw \\
      & & & & & & & & & \\
      & & \vdots & \vdots & & \vdots & \vdots &  & & \\
      & & & & & & & & & \\
      & \lstick{\ket{1}} & \gate{H} & \qw & \ghost{CPS} & \qw & \targ \qwx[-1] & \qw & \gate{H} & \qw
      \gategroup{1}{4}{7}{7}{.7em}{--}
    }
    \]
    \parbox{7cm}{\caption{\label{fig:grover_iter}One iteration of
    Grover's search algorithm with an ancillary qubit used by the
    oracle. $CPS$ is the conditional phase shift operator, while the
    boxed portion is the Grover iteration operator.}}
  \end{center}
\vspace{-7mm}
\end{figure}
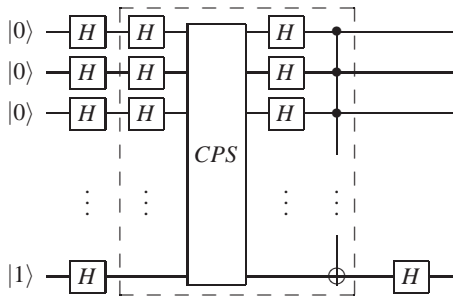

The first benchmark considered is a single iteration of Grover's
quantum search algorithm \cite{Grover1997}, which is depicted in
Figure \ref{fig:grover_iter}. The oracle searches for the last item in
the database \cite{ViamontesEtAl2003}. One iteration is sufficient to
test the effectiveness of the algorithms since the state vector QuIDD
remains isomorphic across all iterations \cite{ViamontesEtAl2003}.

\begin{figure*}[tb]
  \begin{center}
    \begin{tabular}{cc}
      \includegraphics[width=6cm]{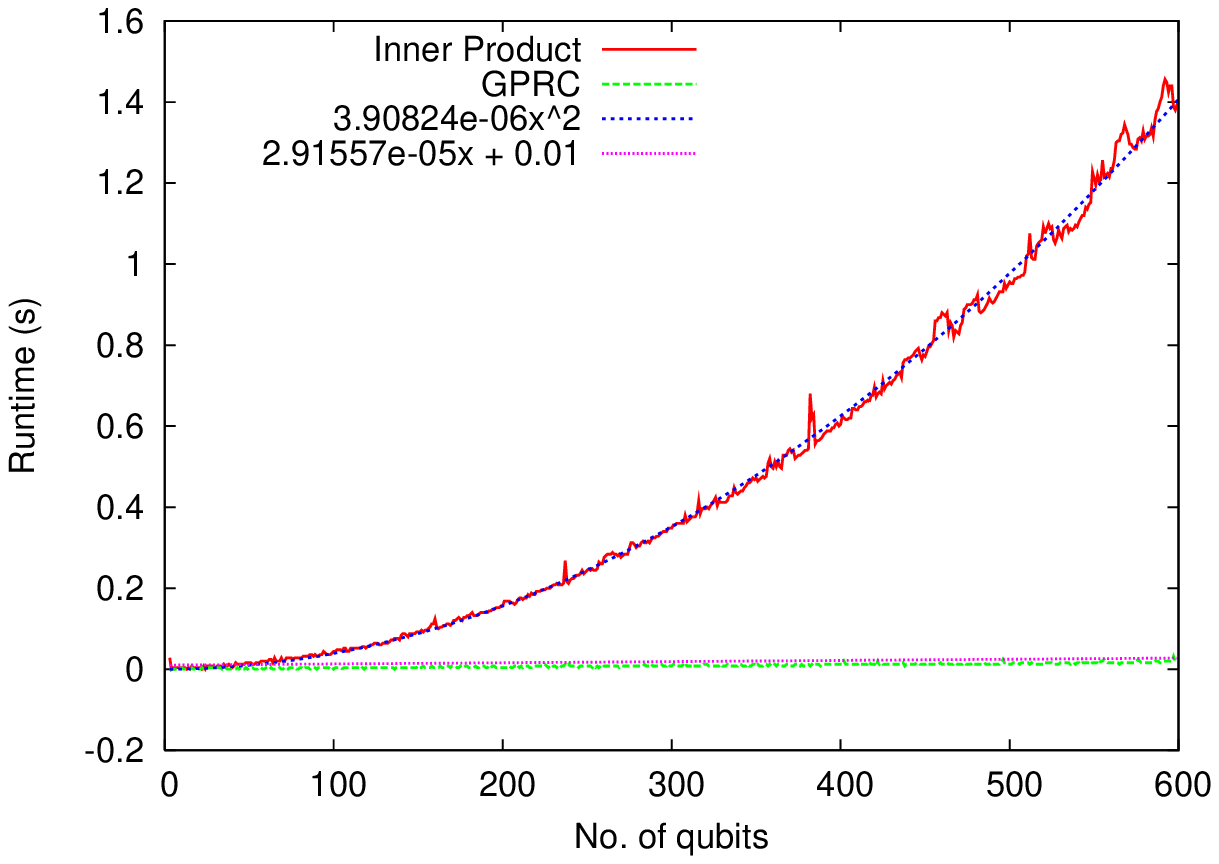} &
      \includegraphics[width=6cm]{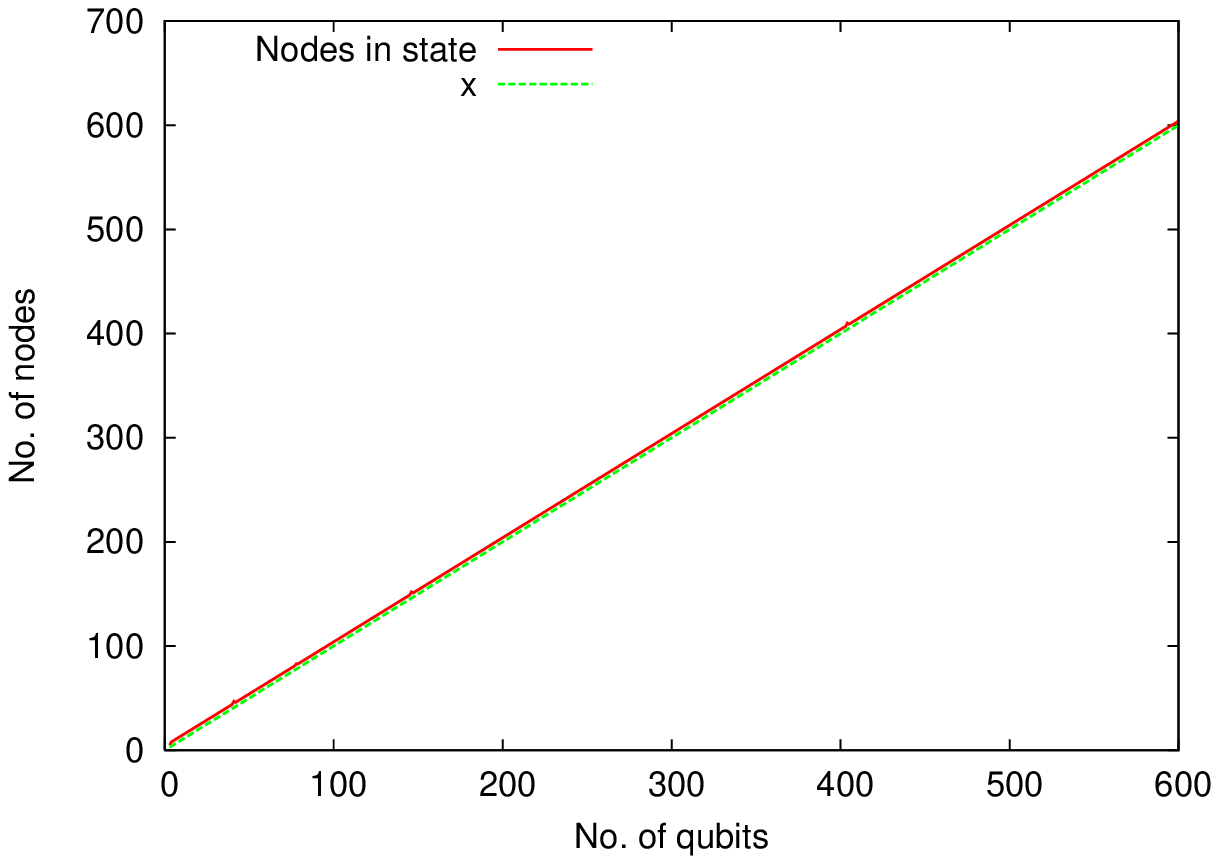} \\
      (a) & (b) \\
      \vspace{-6mm}
    \end{tabular}
    \parbox{16cm}{\caption{\label{fig:grover_state_gp}(a) Runtime
  results and regressions for the inner product and $\mathbf{GPRC}$ on
  checking global-phase equivalence of states generated by a Grover
  iteration. (b) Size in node count and regression of the QuIDD state
  vector.}}
  \end{center}
\vspace{-6mm}
\end{figure*}

Figure \ref{fig:grover_state_gp}a shows the runtime results for the
inner product and $\mathbf{GPRC}$ algorithms (no results are given for
the node-count check algorithm since it runs in $O(1)$ time). The
results confirm the asymptotic complexity differences between the
algorithms. The number of nodes in the QuIDD state vector after a
Grover iteration is $O(n)$ \cite{ViamontesEtAl2003}, which is
confirmed in Figure \ref{fig:grover_state_gp}b. As a result, the
runtime complexity of the inner product should be $O(n^2)$, which is
confirmed by a regression plot within $1\%$ error. By contrast, the
runtime complexity of the $\mathbf{GPRC}$ algorithm should be $O(n)$,
which is also confirmed by another regression plot within $1\%$ error.

\begin{figure*}[tb]
  \begin{center}
    \begin{tabular}{cc}
      \includegraphics[width=6cm]{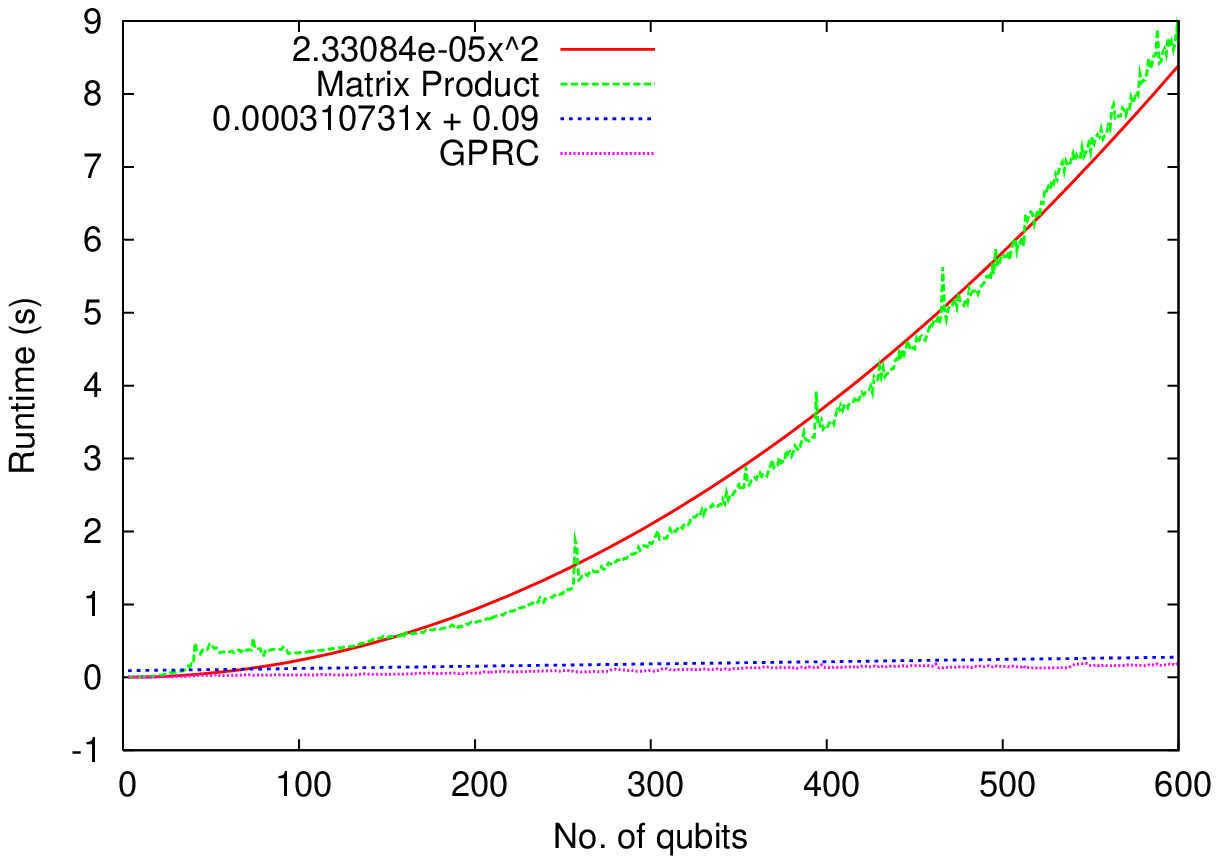} &
      \includegraphics[width=6cm]{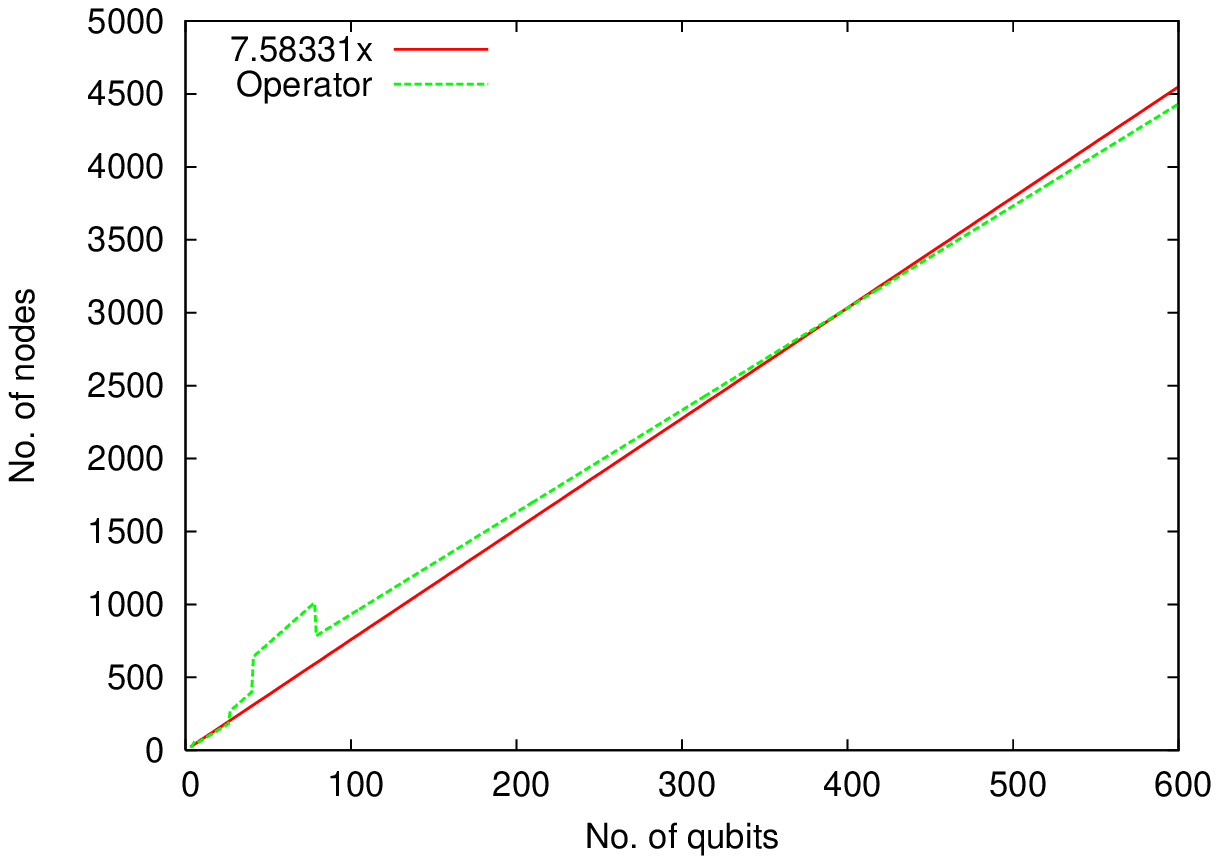} \\
      (a) & (b) \\
      \vspace{-6mm}
    \end{tabular}
    \parbox{16cm}{\caption{\label{fig:grover_op_gp}(a) Runtime results
  and regressions for the matrix product and $\mathbf{GPRC}$ on
  checking global-phase equivalence of the Grover iteration
  operator. (b) Size in node count and regression of the QuIDD
  representation of the operator.}}
  \end{center}
\vspace{-6mm}
\end{figure*}

Figure \ref{fig:grover_op_gp}a shows runtime results for the matrix
product and $\mathbf{GPRC}$ algorithms checking the Grover
operator. Like the state vector, it has been shown that the QuIDD for
this operator grows in size as $O(n)$ \cite{ViamontesEtAl2003}, which
is confirmed in Figure \ref{fig:grover_op_gp}b. Therefore, the runtime
of the matrix product should be quadratic in $n$ but linear in $n$ for
$\mathbf{GPRC}$. Regression plots verify these complexities within
$0.3\%$ error.

\begin{figure}[tb]
  \begin{center}
    \begin{tabular}{@{}r@{}@{}l@{}}
      \begin{tabular}{@{}r@{}}\raisebox{41.5mm}[0pt]{${\scriptstyle x}\left[\begin{array}{l}\\ \\ \\ \\ \\ \\ \end{array}\right.$} \\ 
	  \raisebox{18.5mm}[0pt]{${\scriptstyle 7^x mod 15}\left[\begin{array}{l}\\ \\ \\ \\ \\ \\ \end{array}\right.$}
      \end{tabular} &
      \includegraphics[width=6cm]{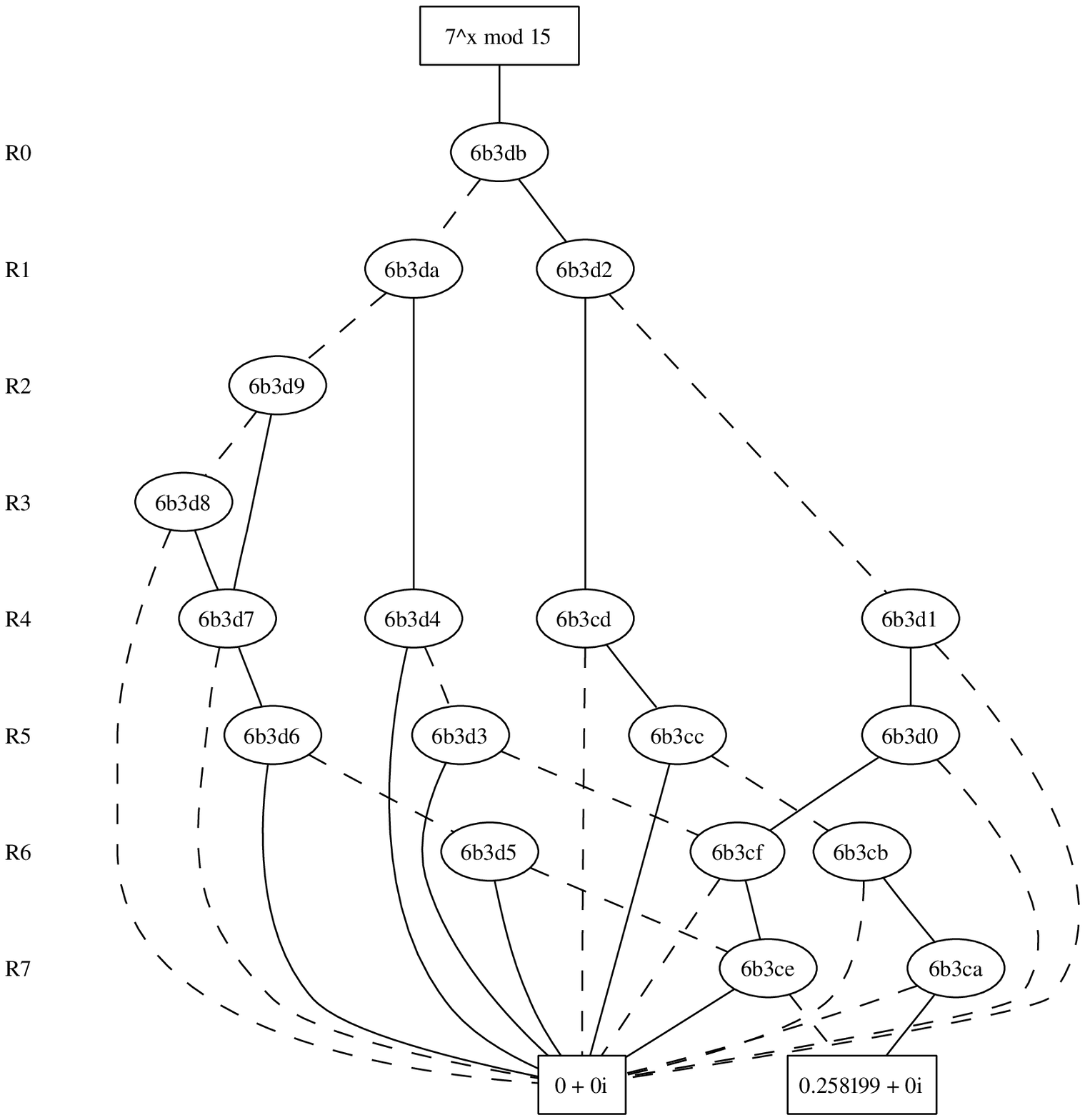}
    \end{tabular}
    \parbox{7.5cm}{\caption{\label{fig:shor_state}A QuIDD state
  combining $x$ and $7^x mod 15$ in binary. The first qubit of each
  partition is least-significant.}}
  \end{center}
\vspace{-6mm}
\end{figure}

The next benchmark compares states in Shor's integer factorization
algorithm \cite{Shor1997}. Specifically, we consider states created by
the modular exponentiation sub-circuit that represent all possible
combinations of $x$ and $f(x, N) = a^x mod N$, where $N$ is the
integer to be factored \cite{Shor1997} (see Figure
\ref{fig:shor_state}). Each of the $O(2^n)$ paths to a non-$0$
terminal represents a binary value for $x$ and $f(x, N)$. Thus, this
benchmark tests performance with exponentially-growing QuIDDs.

Tables \ref{tab:shor_states}a-d show the results of the inner product
and $\mathbf{GPRC}$ for this benchmark. Each $N$ is an integer whose
two non-trivial factors are prime.\footnote{Such integers are likely
to be the ones input to Shor's algorithm since they are the foundation
of modern public key cryptography \cite{Shor1997}.} $a$ is set to $N -
2$ since it may be chosen randomly from the range $[2 . . N - 2]$. In
the case of Table \ref{tab:shor_states}a, states $\ket{\psi}$ and
$\ket{\varphi}$ are equal up to global phase. The node counts for both
states are equal as predicted by Lemma
\ref{lemma:scalar_iso}. Interestingly, both algorithms exhibit nearly
the same performance. Tables \ref{tab:shor_states}b,
\ref{tab:shor_states}c and \ref{tab:shor_states}d contain results for
the cases in which Hadamard gates are applied to the first, middle,
and last qubits, respectively, of $\ket{\varphi}$. The results show
that early termination in $\mathbf{GPRC}$ can enhance performance by
factors of roughly 1.5x to 10x.

\begin{table*}[tb]
  \begin{center}
    \begin{tabular}{cc}
      \scriptsize
      \begin{tabular}{|c|c|c|c|c|c||c|} \hline
	No. of & & Creation & No. of & No. of & Inner Product & GPRC \\
	Qubits & \raisebox{1.5ex}[0pt]{N} & Time (s) & Nodes $\ket{\psi}$ & Nodes $\ket{\varphi}$ & Runtime (s) & Runtime (s) \\ \hline
	$12$ & $4031$ & $11.9$ & $9391$ & $9391$ & $0.30$ & $0.26$ \\ \hline
	$13$ & $6973$ & $24.8$ & $10680$ & $10680$ & $0.34$ & $0.28$ \\ \hline
	$14$ & $12127$ & $55.1$ & $18236$ & $18236$ & $0.54$ & $0.46$ \\ \hline
	$15$ & $19093$ & $128.3$ & $12766$ & $12766$ & $0.41$ & $0.32$ \\ \hline
	$16$ & $50501$ & $934.1$ & $51326$ & $51326$ & $1.7$ & $1.6$ \\ \hline
	$17$ & $69707$ & $1969$ & $26417$ & $26417$ & $0.87$ & $0.78$ \\ \hline
	$18$ & $163507$ & $12788$ & $458064$ & $458064$ & $19.6$ & $19.6$ \\ \hline
	$19$ & $387929$ & $93547$ & $182579$ & $182579$ & $6.62$ & $6.02$ \\ \hline
      \end{tabular} &
      \scriptsize
      \begin{tabular}{|c|c||c|} \hline
	No. of & Inner Product & GPRC \\
	Nodes $\ket{\varphi}$ & Runtime (s) & Runtime (s) \\ \hline
	$10969$ & $0.27$ & $0.036$ \\ \hline
	$11649$ & $0.31$ & $0.036$ \\ \hline
	$19978$ & $0.54$ & $0.06$ \\ \hline
	$13446$ & $0.41$ & $0.036$ \\ \hline
	$55447$ & $1.53$ & $0.2$ \\ \hline
	$27797$ & $0.78$ & $0.084$ \\ \hline
	$521725$ & $19.0$ & $9.18$ \\ \hline
	$194964$ & $6.44$ & $4.40$ \\ \hline
      \end{tabular} \\
      (a) & (b) \\
      \scriptsize
      \begin{tabular}{|c|c|c|c|c|c||c|} \hline
	No. of & & Creation & No. of & No. of & Inner Product & GPRC \\
	Qubits & \raisebox{1.5ex}[0pt]{N} & Time (s) & Nodes $\ket{\psi}$ & Nodes $\ket{\varphi}$ & Runtime (s) & Runtime (s) \\ \hline
	$12$ & $4031$ & $11.9$ & $9391$ & $11773$ & $0.27$ & $0.076$ \\ \hline
	$13$ & $6973$ & $24.8$ & $10680$ & $16431$ & $0.43$ & $0.14$ \\ \hline
	$14$ & $12127$ & $55.1$ & $18236$ & $29584$ & $0.65$ & $0.22$ \\ \hline
	$15$ & $19093$ & $128.3$ & $12766$ & $19207$ & $0.56$ & $0.20$ \\ \hline
	$16$ & $50501$ & $934.1$ & $51326$ & $71062$ & $1.76$ & $0.84$ \\ \hline
	$17$ & $69707$ & $1969$ & $26417$ & $46942$ & $1.24$ & $0.55$ \\ \hline
	$18$ & $163507$ & $12788$ & $458064$ & $653048$ & $31.7$ & $26.1$ \\ \hline
	$19$ & $387929$ & $93547$ & $182579$ & $312626$ & $9.33$ & $6.44$ \\ \hline
      \end{tabular} &
      \scriptsize
      \begin{tabular}{|c|c||c|} \hline
	No. of & Inner Product & GPRC \\
	Nodes $\ket{\varphi}$ & Runtime (s) & Runtime (s) \\ \hline
	$14092$ & $0.21$ & $0.088$ \\ \hline
	$16431$ & $0.27$ & $0.084$ \\ \hline
	$29584$ & $0.53$ & $0.13$ \\ \hline
	$19207$ & $0.50$ & $0.084$ \\ \hline
	$74919$ & $1.51$ & $0.66$ \\ \hline
        $46942$ & $1.13$ & $0.25$ \\ \hline
	$629533$ & $29.6$ & $23.7$ \\ \hline
	$312626$ & $13.0$ & $8.62$ \\ \hline
      \end{tabular} \\
      (c) & (d) \\
    \end{tabular}
    \parbox{16cm}{\caption{\label{tab:shor_states}Performance results
    for the inner product and $\mathbf{GPRC}$ algorithms on checking
    global-phase equivalence of modular exponentiation states. In (a),
    $\ket{\psi} = \ket{\varphi}$ up to global phase. In (b), (c), and
    (d), Hadamard gates are applied to the first, middle, and last
    qubits, respectively, of $\ket{\varphi}$ so that $\ket{\psi} \ne
    \ket{\varphi}$ up to global phase.}}
  \end{center}
\vspace{-4mm}
\end{table*}

In almost every case, both algorithms represent far less than $1\%$ of
the total runtime. Thus, checking for global-phase equivalence among
QuIDD states appears to be an easily achievable task once the
representations are created. An interesting side note is that some
modular exponentiation QuIDD states with more qubits can have more
exploitable structure than those with fewer qubits. For instance, the
$N=387929$ ($19$ qubits) QuIDD has fewer than half the nodes of the $N
= 163507$ ($18$ qubits) QuIDD.

Table \ref{tab:qft_gp} contains results for the matrix product and
$\mathbf{GPRC}$ algorithm checking the inverse Quantum Fourier
Transform (QFT) operator. The inverse QFT is a key operator in Shor's
algorithm \cite{Shor1997}, and it has been previously shown that its
$n$-qubit QuIDD representation grows as $O(2^{2n})$
\cite{ViamontesEtAl2003}. In this case, the asymptotic differences in
the matrix product and $\mathbf{GPRC}$ are very noticeable. Also, the
memory usage indicates that the matrix product may need asymptotically
more intermediate memory despite operating on QuIDDs with the same
number of nodes as $\mathbf{GPRC}$.

\begin{table}[tb]
  \begin{center}
    \scriptsize
    \begin{tabular}{|c|c|c||c|c|} \hline
    No. of & \multicolumn{2}{|c||}{Matrix Product} & \multicolumn{2}{|c|}{GPRC} \\ \cline{2-5}
    Qubits & Time (s) & Mem (MB) & Time (s) & Mem (MB) \\ \hline
    5 & 2.53 & 1.41 & 0.064 & 0.25 \\ \hline
    6 & 22.55 & 6.90 & 0.24 & 0.66 \\ \hline
    7 & 271.62 & 46.14 & 0.98 & 2.03 \\ \hline
    8 & 3637.14 & 306.69 & 4.97 & 7.02 \\ \hline
    9 & 22717 & 1800.42 & 17.19 & 26.48 \\ \hline
    10 & --- & $>2GB$ & 75.38 & 102.4 \\ \hline
    11 & --- & $>2GB$ & 401.34 & 403.9 \\ \hline
    \end{tabular}
    \parbox{7cm} {\caption{\label{tab:qft_gp} Performance results for
    the matrix product and $\mathbf{GPRC}$ algorithms on checking
    global-phase equivalence of the QFT operator used in Shor's
    factoring algorithm. $>2GB$ indicates that a memory usage cutoff
    of 2GB was exceeded.}}
  \end{center}
\vspace{-6mm}
\end{table}

\section{Checking Equivalence up to Relative Phase}
\label{sec:relative_phase}

The relative-phase checking problem can also be solved in many
ways. The first three algorithms are adapted from linear algebra to
QuIDDs, while the last two exploit DD properties directly, offering
asymptotic improvements.

\subsection{Modulus and Inner Product}
\label{sec:mod_inner_product}

Consider two state vectors $\ket{\psi}$ and $\ket{\varphi}$ that are
equal up to relative phase and have complex-valued elements $w_j$ and
$v_k$, respectively, where $j, k = 0, 1, \ldots, N - 1$. Computing
$\ket{\varphi '} = \Sigma_{i = 0}^{N - 1} |v_j| \ket{j}$ and
$\ket{\psi '} = \Sigma_{k = 0}^{N - 1} |w_k| \ket{k} = \Sigma_{k =
0}^{N - 1} |e^{i \theta_k} v_k| \ket{k}$ sets each phase factor to a
$1$, allowing the inner product to be applied as in Subsection
\ref{sec:inner_product}. The complex modulus operations are computed
as $C = \mathbf{Apply}(A, |\cdot|)$ and $D = \mathbf{Apply}(B,
|\cdot|)$ with runtime and memory complexity $O(|A| + |B|)$, which is
dominated by the $O(|A||B|)$ inner product complexity.


\subsection{Modulus and Matrix Product}
\label{sec:mod_matrix_product}

For operator equivalence up to relative phase, two cases are
considered, namely the diagonal relative-phase matrix appearing on the
left or right side of one of the operators. Consider two operators $U$
and $V$ with elements $u_{j, k}$ and $v_{j, k}$, respectively, where
$j, k = 0, \ldots N - 1$. The two cases in which the relative-phase
factors appear on either side of $V$ are described as $u_{j, k} = e^{i
\theta _j} v_{j, k}$ (left side) and $u_{j, k} = e^{i \theta _k} v_{j,
k}$ (right side). In either case the the matrix product check
discussed in Subsection \ref{sec:matrix_product} may be extended by
computing the complex modulus without increasing the overall
complexity. Note that neither this algorithm nor the modulus and inner
product algorithm calculate the relative-phase factors.

\vspace{-2mm}
\subsection{Element-wise Division}
\label{sec:elem_wise_div}

Given the states discussed in Subsection \ref{sec:mod_inner_product},
$w_k = e^{i \theta_k} v_k$, the operation $w_k / v_j$ for each $j = k$
is a relative-phase factor, $e^{i \theta _k}$. The condition $|w_k /
v_j| = 1$ is used to check if each division yields a relative
phase. If this condition is satisfied for all divisions, the states
are equal up to relative phase.

The QuIDD implementation for states is simply $C = \mathbf{Apply}(A,
B, /)$, where $\mathbf{Apply}$ is augmented to avoid division by $0$
and instead return $1$ when two terminal values being compared equal
$0$ and return $0$ otherwise. $\mathbf{Apply}$ can be further
augmented to terminate early when $|w_j / v_i| \ne 1$. $C$ is a QuIDD
vector containing the relative-phase factors. If $C$ contains a
terminal value of $0$, then $A$ and $B$ do not differ by relative
phase. Since a call to $\mathbf{Apply}$ implements this algorithm, the
runtime and memory complexity are $O(|A||B|)$.

Element-wise division for operators is more complicated. For QuIDD
operators $U$ and $V$, $W = \mathbf{Apply}(U, V, /)$ is a QuIDD matrix
with the relative-phase factor $e^{i \theta _j}$ along row $j$ in the
case of phases appearing on the left side and along column $j$ in the
case of phases appearing on the right side. In the first case, all
rows of $W$ are identical, meaning that the support of $W$ does not
contain any row variables. Similarly, in the second case the support
of $W$ does not contain any column variables. A complication arises
when $0$ values appear in either operator. In such cases, the support
of $W$ may contain both variable types, but the operators may in fact
be equal up to relative phase. Figure \ref{fig:rp_div} presents an
algorithm based on $\mathbf{Apply}$ which accounts for these special
cases by using a sentinel value of $2$ to mark valid $0$ entries that
do not affect relative-phase equivalence.\footnote{Any sentinel value
larger than $1$ may be used since such values do not appear in the
context of quantum circuits.} These entries are recursively ignored by
skipping either row or column variables with sentinel children ($S$
specifies row or column variables), which effectively fills copies of
neighboring row or column phase values in their place in $W$. The
algorithm must be run twice, once for each variable type. The size of
$W$ is $O(|U||V|)$ since it is created with a variant of
$\mathbf{Apply}$.

\begin{figure}[tb]
  \begin{center}
    \begin{tabular}{c}
      \framebox[7.5cm][l]{
      \parbox{8cm}{
      \footnotesize
	$\begin{array}{l}
	  \mathbf{RP\_DIV}(A, B, S)\ \{ \\
	  \hspace{1em}\mathbf{if}\ (A\ ==\ New\_Terminal(0))\ \{ \\
	  \hspace{2em}\mathbf{if}\ (B\ !=\ New\_Terminal(0)) \\
	  \hspace{3em}\mathbf{return}\ New\_Terminal(0); \\
	  \hspace{2em}\mathbf{return}\ New\_Terminal(2); \\
	  \hspace{1em}\} \\
	  \hspace{1em}\mathbf{if}\ (Is\_Constant(A)\ \mathbf{and}\ Is\_Constant(B))\ \{ \\
	  \hspace{2em}nrp = Value(A)/Value(B); \\
	  \hspace{2em}\mathbf{if}\ (\mathbf{sqrt}(\mathbf{real}(nrp)*\mathbf{real}(nrp) +\\
	  \hspace{3em}\mathbf{imag}(nrp)*\mathbf{imag}(nrp))\ !=\ 1) \\
	  \hspace{4em}\mathbf{return}\ New\_Terminal(0); \\
	  \hspace{2em}\mathbf{return}\ New\_Terminal(nrp); \\
	  \hspace{1em}\} \\
	  \hspace{1em}\mathbf{if}\ (Table\_Lookup(R, RP\_DIV, A, B, S))\ \mathbf{return}\ R; \\
	  \hspace{1em}v\ =\ Top\_Var(A, B); \\
	  \hspace{1em}T\ =\ \mathbf{RP\_DIV}(A_v, B_v, S); \\
	  \hspace{1em}E\ =\ \mathbf{RP\_DIV}(A_{v'}, B_{v'}, S); \\
	  \hspace{1em}\mathbf{if}\ ((T\ ==\ New\_Terminal(0))\ \mathbf{or} \\
	  \hspace{2em}(E\ ==\ New\_Terminal(0))) \\
	  \hspace{3em}\mathbf{return}\ New\_Terminal(0); \\
	  \hspace{1em}\mathbf{if}\ ((T\ !=\ E)\ \mathbf{and}\ (Type(v)\ ==\ S))\ \{ \\
	  \hspace{2em}\mathbf{if}\ (Is\_Constant(T)\ \mathbf{and}\ Value(T)\ ==\ 2) \\
	  \hspace{3em}\mathbf{return}\ E; \\
	  \hspace{2em}\mathbf{if}\ (Is\_Constant(E)\ \mathbf{and}\ Value(E)\ ==\ 2) \\
	  \hspace{3em}\mathbf{return}\ T; \\
	  \hspace{2em}\mathbf{return}\ New\_Terminal(0); \\
	  \hspace{1em}\} \\
	  \hspace{1em}\mathbf{if}\ (Is\_Constant(T)\ \mathbf{and}\ Value(T)\ ==\ 2) \\
	  \hspace{2em}T\ =\ New\_Terminal(1); \\
	  \hspace{1em}\mathbf{if}\ (Is\_Constant(E)\ \mathbf{and}\ Value(E)\ ==\ 2) \\
	  \hspace{2em}E\ =\ New\_Terminal(1); \\
	  \hspace{1em}R\ =\ \mathbf{ITE}(v, T, E); \\
	  \hspace{1em}Table\_Insert(R, RP\_DIV, A, B, S); \\
	  \hspace{1em}\mathbf{return}\ R; \\
	  \}
	\end{array}
	$
      }}
    \end{tabular}
    \parbox{7.5cm}{\caption{\label{fig:rp_div}Element-wise division
    algorithm.}}
  \end{center}
\vspace{-8mm}
\end{figure}	  

\vspace{-2mm}
\subsection{Non-0 Terminal Merge}
\label{sec:non_zero}

A necessary condition for relative-phase equivalence is that
zero-valued elements of each state vector appear in the same
locations, as expressed by the following lemma.

\begin{lemma}
  \label{lemma:zero_loc}
  A necessary but not sufficient condition for two states
  $\ket{\varphi} = \Sigma_{j = 0}^{N - 1} v_j \ket{j}$ and $\ket{\psi}
  = \Sigma_{k = 0}^{N - 1} w_k \ket{k}$ to be equal up to relative
  phase is that $\forall v_j = w_k = 0$, $j = k$.
\end{lemma}

\noindent
{\bf Proof.} If $\ket{\psi} = \ket{\varphi}$ up to relative phase,
$\ket{\psi} = \Sigma_{k = 0}^{N - 1} e^{i \theta_k} v_k
\ket{k}$. Since $e^{i \theta _k} \ne 0$ for any $\theta$, if any $w_k
= 0$, then $v_j = 0$ must also be true where $j = k$. A
counter-example proving insufficiency is $\ket{\psi} = (0, 1/\sqrt{3},
1/\sqrt{3}, 1/\sqrt{3})^T$ and $\ket{\varphi} = (0, 1/2, 1/\sqrt{2},
1/2)^T$. \hfill $\Box$

QuIDD canonicity may now be exploited. Let $A$ and $B$ be the QuIDD
representations of the states $\ket{\psi}$ and $\ket{\varphi}$,
respectively. First compute $C = \mathbf{Apply}(A, \lceil | \cdot |
\rceil)$ and $D = \mathbf{Apply}(B, \lceil | \cdot | \rceil)$, which
converts every non-zero terminal value of $A$ and $B$ into a
$1$. Since $C$ and $D$ have only two terminal values, $0$ and $1$,
checking if $C = D$ satisfies Lemma \ref{lemma:zero_loc}. Canonicity
ensures this check requires $O(1)$ time and memory. The overall
runtime and memory complexity of this algorithm is $O(|A| + |B|)$ due
to the unary $\mathbf{Apply}$ operations. This algorithm also applies
to operators since Lemma \ref{lemma:zero_loc} also applies to $u_{j,
k} = e^{i \theta _j} v_{j, k}$ (phases on the left) and $u_{j, k} =
e^{i \theta _k} v_{j, k}$ (phases on the right) for operators $U$ and
$V$.


\vspace{-2mm}
\subsection{Modulus and DD Compare}
\label{sec:mod_dd_compare}

A variant of the algorithm presented in Subsection
\ref{sec:mod_inner_product}, which also exploits canonicity, provides
an asymptotic improvement for checking a necessary and sufficient
condition of relative-phase equivalence of states and operators. As in
Subsection \ref{sec:mod_inner_product}, compute $C = \mathbf{Apply}(A,
|\cdot|)$ and $D = \mathbf{Apply}(B, |\cdot|)$. If $A$ and $B$ are
equal up to relative phase, then $C = D$ since each phase factor
becomes a $1$. This check requires $O(1)$ time and memory due to
canonicity. Thus, the runtime and memory complexity is dominated by
the unary $\mathbf{Apply}$ operations, giving $O(|A| + |B|)$.

\vspace{-3mm}
\subsection{Empirical Results for Relative-Phase\\Equivalence Algorithms}
\label{sec:rp_results}

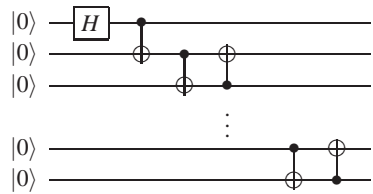
\begin{figure}[tb]
  \begin{center}
    \[
    \Qcircuit @C=1em @R=.01em @!R {
      & \lstick{\ket{0}} & \gate{H} & \ctrl{1} & \qw & \qw & \qw & \qw & \qw & \qw \\
      & \lstick{\ket{0}} & \qw & \targ & \ctrl{1} & \targ & \qw & \qw & \qw & \qw \\
      & \lstick{\ket{0}} & \qw & \qw & \targ & \ctrl{-1} & \qw & \qw & \qw & \qw \\
      & & & & & \vdots & & & & \\
      & \lstick{\ket{0}} & \qw & \qw & \qw & \qw & \qw & \ctrl{1} & \targ & \qw \\
      & \lstick{\ket{0}} & \qw & \qw & \qw & \qw & \qw & \targ & \ctrl{-1} & \qw
    }
    \]
    \vspace{-2mm}
    \parbox{7cm}{\caption{\label{fig:remote_epr}Remote EPR-pair
creation between the first and last qubits via nearest-neighbor
interactions.}}
  \end{center}
\vspace{-6mm}
\end{figure}

The first benchmark for the relative-phase equivalence checking
algorithms creates a remote EPR pair, which is an EPR pair between the
first and last qubits, via nearest-neighbor interactions
\cite{Berman2002}. The circuit is shown in Figure
\ref{fig:remote_epr}. Specifically, it transforms the initial state
$\ket{00 \ldots 0}$ into $(1/\sqrt{2})(\ket{00 \ldots 0} + \ket{10
\ldots 1})$. The circuit size is varied, and the final state is
compared to the state $(e^{0.345i}/\sqrt{2})\ket{00 \ldots 0} +
(e^{0.457i}/\sqrt{2})\ket{10 \ldots 1}$.

\begin{figure*}[tb]
  \begin{center}
    \begin{tabular}{cc}
      \includegraphics[width=6cm]{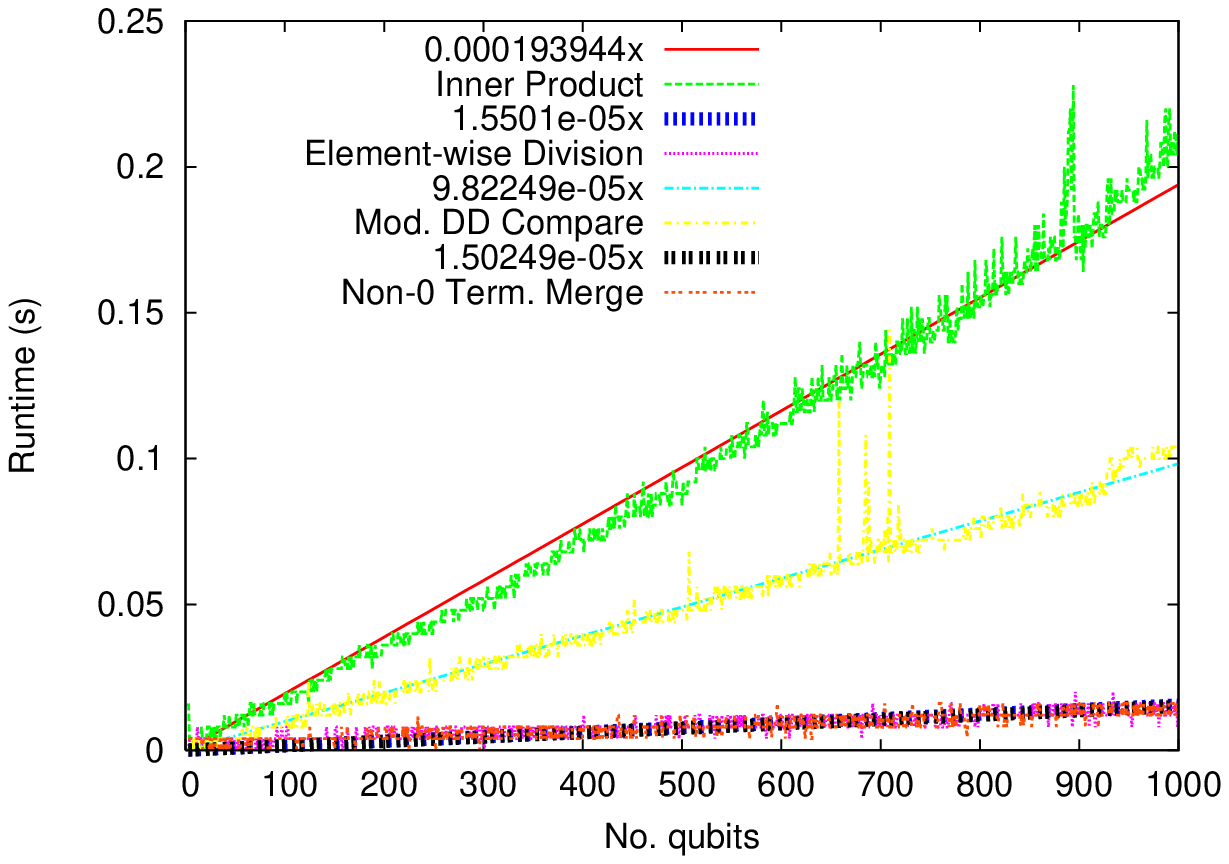} &
      \includegraphics[width=6cm]{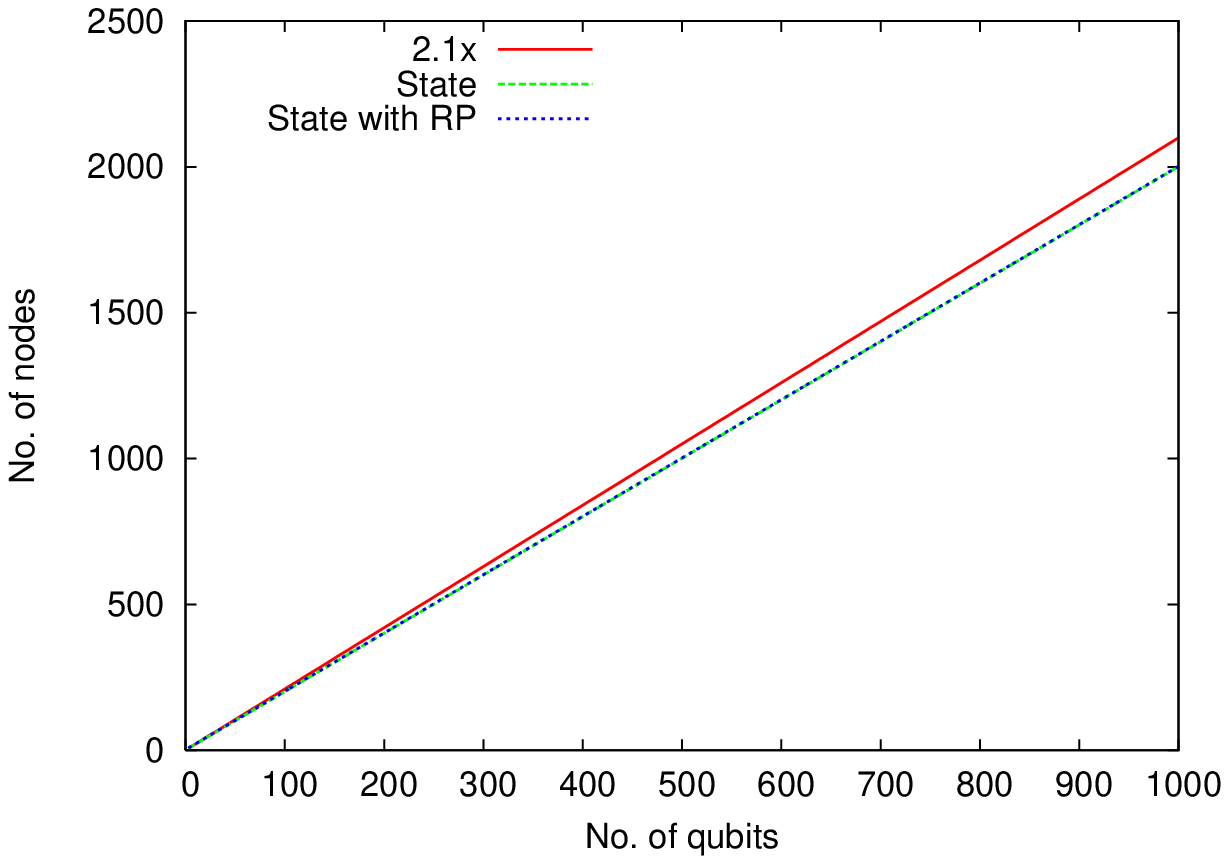} \\
      (a) & (b)
      \vspace{-2mm}
    \end{tabular}
  \parbox{16cm}{\caption{\label{fig:teleport_state_rp}(a) Runtime
  results and (b) size in nodes plotted with regressions for inner
  product, element-wise division, modulus and DD compare, and non-$0$
  terminal merge checking relative-phase equivalence of the remote EPR
  pair circuit.}}
  \end{center}
\vspace{-6mm}
\end{figure*}

The results in Figure \ref{fig:teleport_state_rp}a show that all
algorithms run quickly. The inner product is the slowest, yet it runs
in approximately 0.2 seconds at $1000$ qubits, a small fraction of the
7.6 seconds required to create the QuIDD state vectors. Regressions of
the runtime and memory data reveal linear complexity for all
algorithms to within $1\%$ error. This is not unexpected since the
QuIDD representations of the states grow linearly with the number of
qubits (see Figure \ref{fig:teleport_state_rp}b), and the complex
modulus reduces the number of different terminals prior to computing
the inner product. These results illustrate that in practice, the
inner product and element-wise division algorithms can perform better
than their worst-case complexity. Element-wise division should be
preferred when QuIDD states are compact since unlike the other
algorithms, it computes the relative-phase factors.

\begin{figure}[tb]
  \begin{center}
    \[
    \Qcircuit @C=1em @R=.0005em @!R {
      & & \ctrl{1} & \qw & \qw & \qw & \qw & \qw & \ctrl{1} & \qw & \\
      & & & & & & & & & & \\
      & & \vdots & & & & & & \vdots & & \\
      & & & & & & & & & & \\
      & & \qw & \ctrl{2} & \qw & \qw & \qw & \ctrl{2} & \qw & \qw & \\
      & & \qw & \qw & \ctrl{1} & \qw & \ctrl{1} & \qw & \qw & \qw & \\
      & \lstick{\ket{0}} & \targ \qwx[-3] & \targ & \targ & \gate{e^{-i \Delta t Z}} & \targ & \targ & \targ \qwx[-3] & \qw & \rstick{\ket{0}}
    }
    \]
    \vspace{-2mm}
    \parbox{7cm}{\caption{\label{fig:ham_ckt}A quantum-circuit
    realization of a Hamiltonian consisting of Pauli operators.}}
  \end{center}
\vspace{-3mm}
\end{figure}
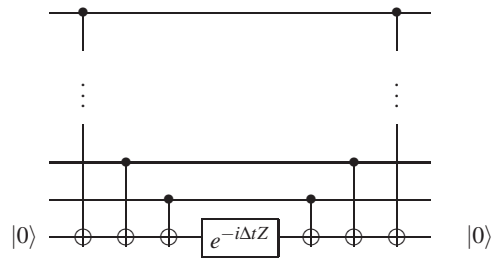

\begin{figure*}[!htb]
  \begin{center}
    \begin{tabular}{cc}
      \includegraphics[width=6cm]{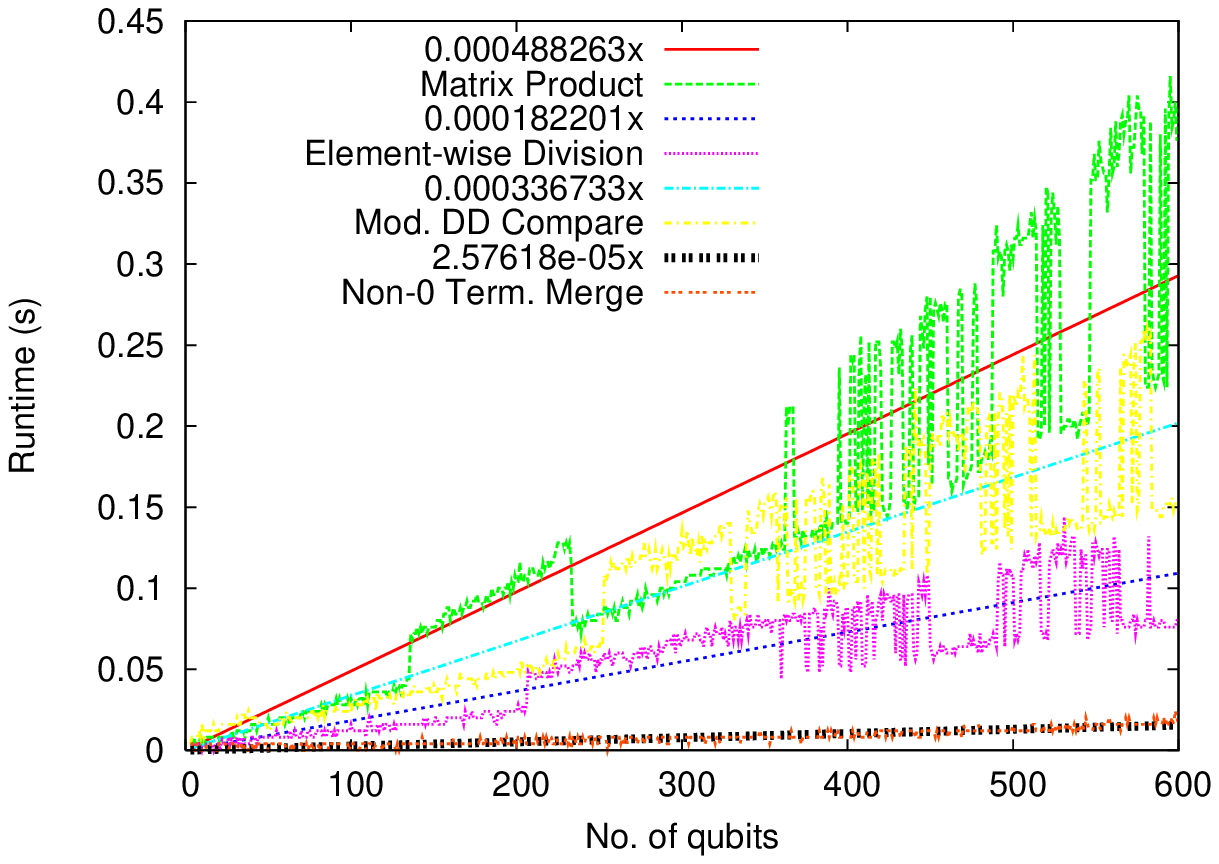} &
      \includegraphics[width=6cm]{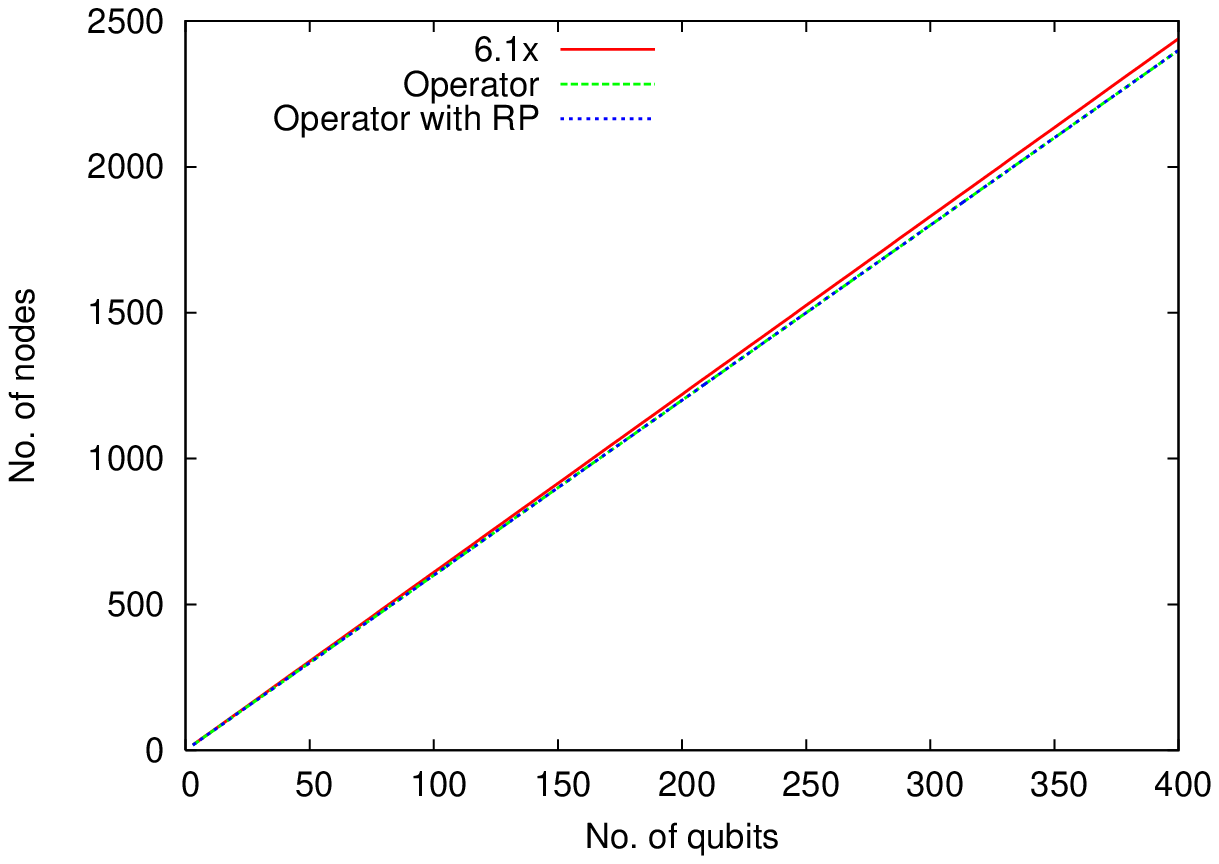} \\
      (a) & (b) \\
      \vspace{-6mm}
    \end{tabular}
    \vspace{-5mm}
  \parbox{16cm}{\caption{\label{fig:hamiltonian_rp}(a) Runtime results
  and (b) size in nodes plotted with regressions for matrix product,
  element-wise division, modulus and DD compare, and non-$0$ terminal
  merge checking relative-phase equivalence of the Hamiltonian $\Delta
  t$ circuit.}}
  \end{center}
\vspace{-1mm}
\end{figure*}

The Hamiltonian simulation circuit shown in Figure \ref{fig:ham_ckt}
is taken from \cite[Figure 4.19, p. 210]{NielsenC2000}.  When its
one-qubit gate (boxed) varies with $\Delta t$, it produces a variety
of diagonal operators, all of which are equivalent up to relative
phase. Empirical results for such equivalence checking are shown in
Figure \ref{fig:hamiltonian_rp}. As before, the matrix product and
element-wise division algorithms perform better than their worst-case
bounds, indicating that element-wise division is the best choice for
compact QuIDDs.



\vspace{-2mm}
\section{Conclusions}
\label{sec:conclusions}

Although DD properties like canonicity enable exact equivalence
checking in $O(1)$ time, we have shown that such properties may be
exploited to develop efficient algorithms for the difficult problem of
equivalence checking up to global and relative phase. In particular,
the global-phase recursive check and element-wise division algorithms
efficiently determine equivalence of states and operators up to global
and relative phase, and compute the phases. In practice, they
outperform QuIDD matrix and inner products, which do not compute
relative-phase factors. Other QuIDD algorithms presented in this work,
such as the node-count check, non-$0$ terminal merge, and modulus and
DD compare, exploit other DD properties to provide even faster checks
but only satisfy necessary equivalence conditions. Thus, they should
be used to aid the more robust algorithms. A summary of the
theoretical results is provided in Table \ref{tab:all_methods}.

\begin{table}[tb]
  \scriptsize
  \begin{center}
    \begin{tabular}{|@{}c@{}|@{}c@{}|@{\ }c@{\ }|l@{}|@{\ }c@{}|@{\ }c@{}|} \hline
      & & &  & $O(\cdot)$ time & $O(\cdot)$ time \\
      Algorithm & \raisebox{1.5ex}[0pt]{Phase} & \raisebox{1.5ex}[0pt]{Finds} & \raisebox{1.5ex}[0pt]{Necessary \&} & complexity: & complexity: \\
      & \raisebox{1.5ex}[0pt]{type} & \raisebox{1.5ex}[0pt]{phases?} & \raisebox{1.5ex}[0pt]{sufficient?} & best-case & worst-case \\ \hline \hline
      Inner & & & & & \\
      Product & \raisebox{1.5ex}[0pt]{Global} & \raisebox{1.5ex}[0pt]{Yes} & \raisebox{1.5ex}[0pt]{N. \& S.} & \raisebox{1.5ex}[0pt]{$|A||B|$} & \raisebox{1.5ex}[0pt]{$|A||B|$} \\ \hline
      Matrix & & & & & \\
      Product & \raisebox{1.5ex}[0pt]{Global} & \raisebox{1.5ex}[0pt]{Yes} & \raisebox{1.5ex}[0pt]{N. \& S.} & \raisebox{1.5ex}[0pt]{$(|A||B|)^2$} & \raisebox{1.5ex}[0pt]{$(|A||B|)^2$} \\ \hline 
      Node-Count & Global & No & N. only & $1$ & $1$ \\ \hline
      {\bf Recursive} & & & & & \\
      {\bf Check} & \raisebox{1.5ex}[0pt]{\bf Global} & \raisebox{1.5ex}[0pt]{\bf Yes} & \raisebox{1.5ex}[0pt]{\bf N. \& S.} & \raisebox{1.5ex}[0pt]{$\mathbf{1}$} & \raisebox{1.5ex}[0pt]{$\mathbf{|A| + |B|}$} \\ \hline \hline
      Modulus and & & & & & \\
      Inner Product & \raisebox{1.5ex}[0pt]{Relative} & \raisebox{1.5ex}[0pt]{No} & \raisebox{1.5ex}[0pt]{N. \& S.} & \raisebox{1.5ex}[0pt]{$|A||B|$} & \raisebox{1.5ex}[0pt]{$|A||B|$} \\ \hline
      {\bf Element-wise} & & & & & \\
      {\bf Division} & \raisebox{1.5ex}[0pt]{\bf Relative} & \raisebox{1.5ex}[0pt]{\bf Yes} & \raisebox{1.5ex}[0pt]{\bf N. \& S.} & \raisebox{1.5ex}[0pt]{$\mathbf{|A||B|}$} & \raisebox{1.5ex}[0pt]{$\mathbf{|A||B|}$} \\ \hline
      Non-$0$ & & & & & \\
      Terminal Merge & \raisebox{1.5ex}[0pt]{Relative} & \raisebox{1.5ex}[0pt]{No} & \raisebox{1.5ex}[0pt]{N. only} & \raisebox{1.5ex}[0pt]{$|A| + |B|$} & \raisebox{1.5ex}[0pt]{$|A| + |B|$} \\ \hline
      Modulus and & & & & & \\
      DD Compare & \raisebox{1.5ex}[0pt]{Relative} & \raisebox{1.5ex}[0pt]{No} & \raisebox{1.5ex}[0pt]{N. \& S.} & \raisebox{1.5ex}[0pt]{$|A| + |B|$} & \raisebox{1.5ex}[0pt]{$|A| + |B|$} \\ \hline
    \end{tabular}
    \parbox{7cm}{\caption{\label{tab:all_methods}Key properties of the
    QuIDD-based phase-equivalence checking algorithms.}}
  \end{center}
\vspace{-6mm}
\end{table}

The algorithms presented here enable QuIDDs and other DD
datastructures to be used in synthesis and verification of quantum
circuits. A fair amount of work has been done on optimal synthesis for
small quantum circuits as well as heuristics for larger circuits via
circuit transformations \cite{PrasadEtAl2007,
ShendeEtAl2006}. Equivalence checking in particular plays a key role
in some of these techniques since it is often necessary to verify the
correctness of the transformations. Future work will determine how
these equivalence checking algorithms may be used as primitives to
enhance such heuristics.

{\bf Acknowledgements. }
This work was funded by the Air Force Research Laboratory.  The views
and conclusions contained herein are those of the authors and should
not be interpreted as necessarily representing official policies or
endorsements of employers and funding agencies.

\end{document}